\documentclass[twocolumn,showpacs,preprintnumbers,amsmath,amssymb]{revtex4}

\usepackage{graphicx}
\usepackage{dcolumn}
\usepackage{bm}

\begin{document}

\preprint{}

\title{Point--like gamma ray sources as signatures \\ 
of distant accelerators of ultra high energy cosmic rays}

\author{Stefano Gabici}
 \email{Stefano.Gabici@mpi-hd.mpg.de}
\author{Felix Aharonian}%
\affiliation{%
Max--Planck--Institut f\"ur Kernphysik,
Saupfercheckweg 1, 69117 Heidelberg, Germany 
}%


\date{\today}

\begin{abstract}
We discuss the possibility of observing distant accelerators of ultra high energy cosmic rays in synchrotron gamma rays.
Protons propagating away from their acceleration sites produce extremely energetic electrons during photo-pion interactions with cosmic microwave background photons.
If the accelerator is embedded in a magnetized region, these electrons will emit high energy synchrotron radiation.
The resulting synchrotron source is expected to be point-like, steady and detectable in the GeV-TeV energy range
if the magnetic field is at the nanoGauss level.
\end{abstract}

\pacs{Valid PACS appear here}

\maketitle

Cosmic rays (CR) are observed up to energies beyond $10^{20} {\rm eV}$. 
Even though it is commonly believed that particles with energies up to $\sim 10^{15} {\rm eV}$ are accelerated in supernova remnants, we are still far from having a clear picture of the origin of higher energy particles. 
In particular, the Ultra High Energy Cosmic Rays (UHECR) observed at the end of the spectrum constitute a real challenge for theoretical models, since their acceleration requires extreme conditions that can hardly be fulfilled by known astrophysical objects.
Previous studies have considered a number of potential sources, including gamma ray bursts, active galactic nuclei, large scale jets and neutron stars, but results are still inconclusive \cite{reviews}.  

Moreover, 
even if some astrophysical objects accelerate UHECRs, a suppression, known as the Greisen-Zatsepin-Kuzmin (GZK) cutoff, should be observed in the CR spectrum at an energy $E_{GZK} \sim 7 \; 10^{19} {\rm eV}$, if UHECR sources are homogeneously distributed in the Universe \cite{GZK}. 
This is because during propagation energetic protons undergo $p\gamma$ interactions (mainly photo-pion production) in the 
Cosmic Microwave Background (CMB) photon field, with an energy loss length which decreases dramatically from $\sim 650$ to $\sim 20 \, {\rm Mpc}$ between energies of $7 \; 10^{19}$ to $3 \; 10^{20} {\rm eV}$ \cite{pgamma}. 
The detection of a number of events above $E_{GZK}$ 
raised interest in non-acceleration models 
that
involve new physics and do not require the existence of a cutoff in the spectrum \cite{reviews}.
However, the issue of the presence or absence of the GZK feature 
is still debated \cite{daniel}. 
In the following we assume that astrophysical objects capable of accelerating UHECRs 
do exist and we discuss a possible way to identify them. 

If UHECRs are 
not appreciably
deflected by the 
intergalactic magnetic field \cite{klaus} (for a different view see \cite{minia}), 
their arrival directions should point back to the actual position of their accelerators.
In principle, this fact could be used to identify accelerators located at a distance smaller than the proton loss length, that constitutes a sort of horizon for CRs.
The interactions between protons and CMB photons, responsible for the existence of this horizon, also generate secondary gamma rays and electron-positron pairs (hereafter we refer to both electrons and positrons as {\it electrons}) that in turn start an electromagnetic cascade in the universal photon background \cite{cascade}.
Low energy electrons produced in the cascade are effectively deflected by the intergalactic magnetic field, and their radiation would appear to a distant observer as an extended gamma ray halo \cite{halo}.
However, if the magnetic field close to the accelerator is at the ${\rm nG}$ level, first generation electrons generated during $p\gamma$ interactions cool rapidly  by emitting ${\rm GeV}$ synchrotron photons and the development of the cascade is strongly inhibited \cite{felixp}. 

We discuss the possibility to detect this synchrotron radiation from UHECR accelerators. 
Such a possibility is of great interest because of the following reasons.
Both the synchrotron emitting electrons and parent protons are extremely energetic and not appreciably deflected by the intergalactic magnetic field, at least on the first Mpc distance scale.
For this reason, the observed radiation is expected to be point-like, and thus easily detectable and distinguishable from the extended cascade component.
Remarkably, a possible detection of these sources would allow us to infer the value of the intergalactic magnetic field close to the accelerator, constraining it in the range between a fraction and a few tens of ${\rm nG}$.
Finally, since the Universe is transparent to ${\rm GeV}$ photons, also powerful UHECR accelerators located outside the CR horizon might be identified in this way.  

{\it Synchrotron gamma rays.--}
Consider a steady and isotropic accelerator of UHECRs.
The actual nature of the accelerator as well as the details of the mechanism through which protons are accelerated up to ultra high energies are irrelevant for the discussion here.
The only assumption we make is that protons with energies up to $10^{20} {\rm eV}$ and beyond 
can leak out of the acceleration region and propagate away. 
In a mildly magnetized environment, such energetic protons  
follow almost rectilinear trajectories up to a distance roughly equal to the proton loss length.
For proton energies greater than $\sim 5 \times 10^{19} {\rm eV}$ the most relevant loss channel is photo-pion production in interactions with CMB photons and the loss length is $\lambda \sim 150 \; {\rm Mpc}$ at $10^{20} {\rm eV}$ and rapidly decreases at higher energies.
This implies that, if the accelerator is located at a distance much greater than $\lambda$, it cannot be directly observed as a source of UHECRs.

During photo-pion interactions both neutral and charged pions are produced. They in turn decay giving as final products photons, electrons and neutrinos.
If $E_{p,20}$ is the energy of the incoming proton in units of $10^{20} {\rm eV}$, the typical energy of the outgoing photons, electrons and neutrinos are $E_{\gamma} \sim 10^{19} E_{p,20} {\rm eV}$ and $E_{e^{\pm}} \sim E_{\nu} \sim 5 \; 10^{18} E_{p,20} {\rm eV}$ respectively \cite{pgamma}. 
Electrons and photons interact via Compton and pair production processes with soft photons in the CMB and radio background.
In general, this would lead to the development of an electromagnetic cascade, in which the number of electrons and photons increases rapidly.
In fact, due to the extremely high energy of the particles considered here, each interaction occurs in the limit $\epsilon_b E \gg 1$, where $\epsilon_b$ and $E$ are the energies of the background photons and of the energetic electron (or photon) respectively, both calculated in units of the electron rest mass energy.
Under this condition the Compton scattering is in the extreme Klein-Nishina limit, namely, the upscattered photon carries away most of the energy of the incoming electron.
The same happens during a pair production event, in which most of the energy goes to one of the two outgoing electrons.
The fraction of the energy lost by the energetic particle is approximately $f \sim 1/ln(2\epsilon_b E)$ \cite{pgamma}, which reduces to only a few percent if we consider a $10^{19} eV$ electron interactiong with a CMB photon.
Therefore the problem reduces essentially to a single-particle problem, in which a leading particle loses continuously energy and changes state from electron to photon and back: $e^{\pm} \rightarrow \gamma \rightarrow e^{\pm}$\dots due to alternate Compton and pair production interactions.
Thus, the effective loss length for an energetic electron can be identified with the loss length of the leading particle \cite{gould}. 

\begin{figure}
\includegraphics[width=0.4\textwidth]{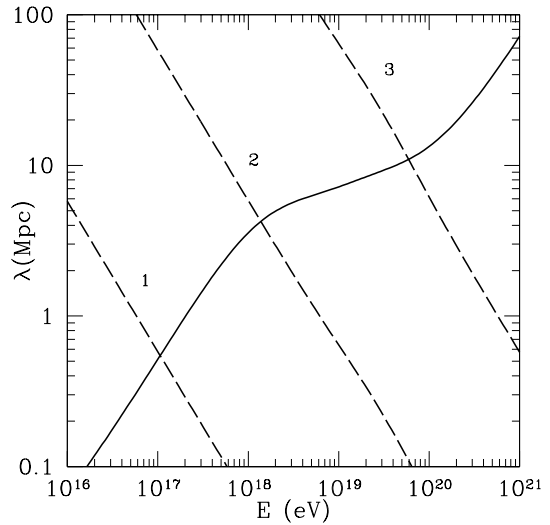}
\caption{\label{eloss}Effective loss length for electrons. Solid line refers to Compton and pair production processes in the CMB and radio background (assumed to have a low frequency cutoff at $2 \sim {\rm MHz}$ \cite{radiobackground}). Dashed lines refer to synchrotron losses in a magnetic field of $10$, $1$ and $0.1 ~ {\rm nG}$ (curves 1, 2 and 3).
}
\end{figure}

If a magnetic field $B$ is present close to the accelerator, electrons also lose energy via synchrotron emission. 
In Fig.~\ref{eloss} we show the effective electron loss length calculated following \cite{gould} (solid line), together with the synchrotron loss length for different values of the magnetic field strength (dashed lines, $0.1$, $1$ and $10 ~ {\rm nG}$ top to bottom).
Above a certain energy $E_{*}$, synchrotron losses 
dominate over Compton/pair production ones.
If we assume that the magnetic field is 1 nG,
then $E_* \sim 10^{18} {\rm eV}$ and the energy of the synchrotron photons falls in the ${\rm GeV}$ range: $E_{syn} \sim 10 \, [B/({\rm nG})] \, [E/(10^{19}{\rm eV})]^2 \, {\rm GeV}$.
This implies that the presence of UHECR accelerators 
could in principle be revealed 
by means of gamma ray observations.

A necessary condition for the production of synchrotron photons is that the size of the magnetized region surrounding the accelerator must be greater than or comparable with the interaction length 
of UHECR protons.
Thus, we consider here a region os size 
$\sim 10 \div 20 ~ {\rm Mpc}$, large enough to allow effective electron production inside the magnetized environment.
Superclusters of galaxies constitute an example of a very large and appreciably magnetized region satisfying our requirement \cite{Bfield}. 
It is important to stress that our results are sensitive only to the value of the magnetic field {\it close} to the UHECR source, while they are unaffected by the value of the field on much larger scales. \\
{\it Angular size.--}
A mild magnetic field close to the source can slightly deflect UHECRs.
Adopting the small scattering-angle approximation and neglecting energy losses , an order of magnitude estimate of the deflection angle for a proton of energy $E$ propagating in a magnetized region of size $d_p$ is: $\vartheta_p \sim 1^o (10^{20}{\rm eV}/E) (B/{\rm nG}) \sqrt{(d_p/20{\rm Mpc})} \sqrt{(l_c/{\rm Mpc})}$, where $l_c$ is the magnetic field coherence length \cite{deflection}.
A fraction of the protons undergoes $p\gamma$ interactions before escaping the magnetized region, producing extremely energetic electrons.
If the magnetic field is at the $nG$ level, electrons with energy in excess of $\sim 10^{18} {\rm eV}$ cool fast via emission of synchrotron gamma rays, depositing all their energy along the direction of their motion.
This implies that an observer at a distance $D$ would see a gamma ray source with angular size $\vartheta_{obs} \le 2 \vartheta_p (d_p/D) \sim 3^{\prime} ({\rm Gpc}/D)$.
This leads to the important conclusion that, even if the gamma rays are produced in an extended region of size $d_p \sim 20 \; {\rm Mpc}$ surrounding the accelerator, the source would appear point-like to a distant observer.
In particular, GLAST, whose angular resolution at energies in the range $1 \div 10 \; {\rm GeV}$ is roughly a fraction of a degree, will classify these sources as point-like if they are
located at a distance greater than $D \sim 50 \div 100 \; {\rm Mpc}$. 
For Imaging Atmospheric Cherenkov Telescope arrays (IACT) operating at energies above $100 \; {\rm GeV}$ this minimum distance is $\sim 1 \; {\rm Gpc}$, since their angular resolution is of the order of a few arcminutes.
Electrons with energy smaller than $E_*$ are not affected by synchrotron losses and start an electromagnetic cascade, that in principle might be detected at ${\rm TeV}$ energies by IACTs 
\cite{carlo}.
However, low energy electrons produced during the last steps of the cascade are effectively isotropized in the intergalactic magnetic field if their Larmor radius is smaller than or comparable with the Compton cooling length.
This condition is satisfied when the magnetic field is greater than $\sim 10^{-12} (E_{\gamma}/{\rm TeV}) \; {\rm G}$, $E_{\gamma}$ being the energy of the Compton-upscattered photons. 
This implies that, unless the intergalactic magnetic field is extremely weak, electrons will emit Compton photons isotropically and the cascade will result in the formation of a very extended, hardly detectable gamma ray halo \cite{halo}.
For this reason, we consider here only the synchrotron component. \\
\begin{figure}
\includegraphics[width=0.5\textwidth]{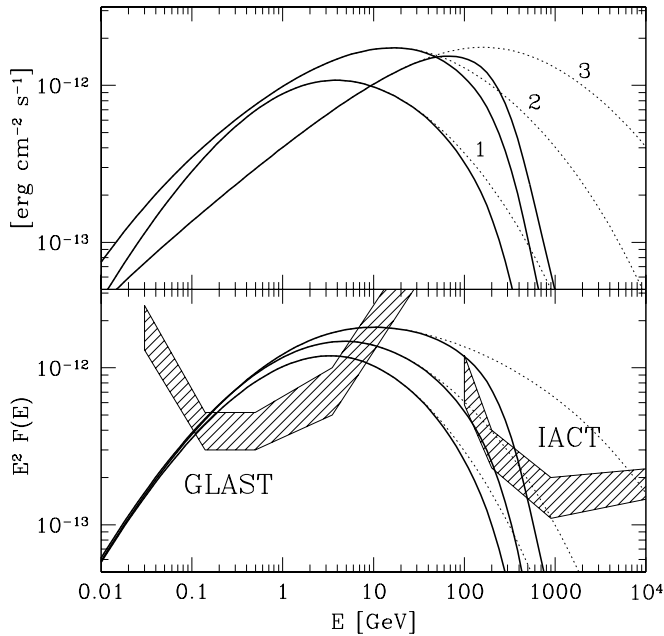}
\caption{\label{spectra}Photon spectra for a source 
at a distance of $1{\rm Gpc}$ 
in a $20{\rm Mpc}$ magnetized region. The luminosity in UHECRs is $10^{46}{\rm erg/s}$, the proton spectral index is $\delta = 2$. TOP: $E_{cut} = 10^{21} {\rm eV}$, magnetic field $0.5$ (curve 1), $5$ (2), $50{\rm nG}$ (3). BOTTOM: $E_{cut} = 5 \, 10^{20}$, $10^{21}$, $5 \, 10^{21} {\rm eV}$, magnetic field is $1{\rm nG}$. Dotted lines represent the intrinsic spectra, solid lines show the effect of absorption in the infrared background. The sensitivities of GLAST and of a generic IACT are shown.}
\end{figure}
{\it Detectability and energetics.--}
Fig.~\ref{spectra} shows synchrotron spectra 
for an UHECR source
at a distance of $1 ~ {\rm Gpc}$.
Steady state proton and electron spectra have been calculated taking into account energy losses (photopion production and Compton/pair-production processes respectively) and proton escape from the magnetized region with a caracteristic time $d_p/c$. The size of the magnetized region is $d_p = 20 {\rm Mpc}$.
Solid lines have been computed 
taking into account the opacity of the Universe to very high energy photons due to pair production in the cosmic infrared background \cite{IRabs}
, while dotted lines show the unabsorbed spectra.
The total luminosity in UHECRs with energy above $10^{19} {\rm eV}$ is $L_{UHE} = 10^{46} {\rm erg/s}$, with a differential energy distribution $Q(E) = Q_0 E^{-\delta} exp(-E/E_{cut})$.
We further assume $\delta =2$, though results are quite insensitive to the slope of the CR spectrum.

In the top panel of Fig.~\ref{spectra} we fix $E_{cut} = 10^{21} {\rm eV}$ and we consider three different values for the magnetic field:  $0.5$, $5$ and $50 \; {\rm nG}$ (curves $1$, $2$ and $3$ respectively).
If the magnetic field is significantly greater than $\sim 50 \; {\rm nG}$, the peak of the emission falls at TeV energies, where absorption is very strong.
The absorbed photons start an electromagnetic cascade that will appear as an extended halo.
On the other hand, if the field is well below $\sim 0.5 \; {\rm nG}$, synchrotron emission becomes unimportant and again the cascade contribution dominates.
However, for the broad interval of values of the magnetic field strength between $0.5$ and $50 \; {\rm nG}$, the formation of a synchrotron point-like gamma ray source seems to be unavoidable. 

In the bottom panel of Fig.~\ref{spectra}, our predictions are compared with the sensitivities of GLAST and of a generic 
IACT array such as HESS \cite{texas}. 
A magnetic field of $1 \; {\rm nG}$ is assumed and the three different curves refer to values of the cutoff energy in the proton spectrum equal to $E_{cut} = 5 \; 10^{21}$, $10^{21}$ and  $10^{21} {\rm eV}$ (top to bottom).
For such a magnetic field, the condition for the detectability of a point source by GLAST is roughly $L_{UHE} \ge 8 \; 10^{43} \div 2 \; 10^{44} (D/100{\rm Mpc})^2 {\rm erg/s}$ for $\delta$ in the range $2.0 \div 2.6$.
In contrast, for IACT arrays the minimum detectable luminosity is about 2 orders of magnitude higher, since the source has to be located at a distance of $\sim 1 {\rm Gpc}$ in order to appear point-like.
Moreover, distant sources may be undetectable above $100 \; {\rm GeV}$ due to the strong absorption in the infrared background.
In this case, the maximum energy of UHECRs is very important, since it determines the extension of the gamma ray spectrum towards high energies.
Since the peak of the emission falls at $\sim 10 ~ {\rm GeV}$, the future IACT arrays operating in the energy range $10 \div 100$ GeV would be powerful tools to search for these sources.  

If the CR spectrum is a smooth power law with index $\delta = 2$ down to GeV energies, the required total CR luminosity for a source to be detected by GLAST is $L_{CR} \ge 5 \; 10^{44} (D/100{\rm Mpc})^2 {\rm erg/s}$.
If CRs 
are beamed along one axis, the luminosity is reduced by a factor $f_b \sim 0.02 (\vartheta_b/10^o)$, $\vartheta_b$ being the beaming angle.
In this case, the detectability condition reads: $L_{CR} \ge 10^{43} (f_b/0.02) (D/100{\rm Mpc})^2 {\rm erg/s}$.
This luminosity is small if compared, for example, with the kinetic power of an AGN jet, that can be as high as 
$\sim 10^{47} {\rm erg/s}$ \cite{ghisella}.
Thus, 
astrophysical object that could in principle satisfy the energy requirement for detectability do exist.

In calculating the spectra shown in Fig.~\ref{spectra} we assumed the UHECR accelerator to be active for a time $t_{ON}$ long enough to reach steady state.
This situation is achieved if the accelerator lifetime is greater than the $p\gamma$ interaction time, namely, if $t_{ON} \ge 100 \; {\rm Myr}$.
Under this assumption, the total energy deposited in CRs during the whole source lifetime is $E_{CR} \ge 3 \; 10^{58} (f_b/0.02) (D/100{\rm Mpc})^2 {\rm erg}$.
If the UHECR accelerator is located inside a rich cluster of galaxies, then the bulk of this nonthermal energy accumulates in it, due to effective diffusive confinement of CRs with energy up to at least $10^7 {\rm GeV}$ in the $\mu {\rm G}$ intracluster magnetic field \cite{confinement}. 
Such an amount of energy can be easily stored in a rich cluster, whose total thermal energy can be as high as $10^{64} {\rm erg}$ \cite{confinement}.

Of course, if the CR spectrum is much steeper than $\delta = 2$, the problem of energetics could be difficult to circumvent, unless a mechanism is found to limit acceleration to ultra high energy particles only \cite{daniel2,carlo} \footnote{If CRs are accelerated only above an energy $E_{min}$ 
then $L_{CR} \propto E_{min}^{2-\delta}/(\delta-2)$. For a steep spectrum $\delta = 2.6$ and $E_{min} = 1 GeV$ the required luminosity is huge $L_{CR} \sim 2.4 \; 10^{50} erg/s$}.

Thus, calculations show that for objects located within several hundreds of megaparsecs and with jets pointed to the observer, GLAST and Cherenkov telescopes like HESS should be able to provide meaningful probes of this radiation. Distant objects beyond 1 Gpc will be possibly detected only by the next generation instruments with significantly improved sensitivities. 

{\it Bursting sources.--}
The results presented above can be qualitatively generalized to the case of a bursting source.
Consider a short burst releasing an energy $E_{UHE}^{burst}$ in the form of UHECRs.
The deflection of UHECRs in magnetic fields results in a time broadening of the pulse over a time roughly equal to the delay time $\delta\tau \sim 5 \, 10^3 (10^{20}{\rm eV}/E)^2 (B/{\rm nG})^2 (d_p/20{\rm Mpc})^2 (l_c/{\rm Mpc}) \, {\rm yr}$ \cite{waxman}.
The duration of the synchrotron 
emission is expected to be roughly the same.
If the intrinsic duration of the burst $t_b$ is longer than $\delta\tau$, time broadening is not important.
To satisfy the detectability condition, the total energy released in UHECRs must be $E_{UHE}^{burst} \ge L_{UHE}^{ss} t_{obs} \sim 2 \; 10^{53} (t_{obs}/\delta\tau) (f_b/0.02) (D/100 {\rm Mpc})^2 {\rm erg/s}$ where $L_{UHE}^{ss}$ is the detectability condition for the steady state situation 
and $t_{obs} \sim \delta\tau + t_b$ 
is the observed duration of the burst.
This is a huge amount of energy to be released in a single explosion, orders of magnitude above the energy $\sim 10^{51} {\rm erg}$ that is believed to be converted into UHECRs during a gamma ray burst \cite{waxman,mario}.
Thus, if UHECRs are produced during short bursts, the related synchrotron emission is far too faint to be detected. 
However, if the accelerator remains active for a time $\gg \delta\tau$ the energy requirement 
is significantly reduced. 

{\it Neutrinos.--}
Muon and electron neutrinos are 
produced during $p\gamma$ interactions in the ratio 2:1.
Due to neutrino oscillation the ratio between the three flavors becomes 1:1:1 at Earth.
For the source considered in Fig.~\ref{spectra} the total neutrino flux is $E_{\nu}^2 F_{\nu} \sim 1 \; {\rm EeV/km^2/yr}$ at energy $5 \; 10^{18} {\rm eV}$.
These energies will be probed by the ANITA and AUGER experiments. The possibility of a detection will depend on their performances in recognizing steady point sources. 
The detection of neutrinos 
could serve as an unequivocal
signature of UHECR acceleration.\\
The authors thank S. Barwick, P. Blasi, P. Coppi, C. Ferrigno, J. Hinton and E. Resconi for useful discussions.
SG acknowledges support from the Humboldt Foundation.


\begin{thebibliography}{99}
\expandafter\ifx\csname natexlab\endcsname\relax\def\natexlab#1{#1}\fi
\expandafter\ifx\csname bibnamefont\endcsname\relax
  \def\bibnamefont#1{#1}\fi
\expandafter\ifx\csname bibfnamefont\endcsname\relax
  \def\bibfnamefont#1{#1}\fi
\expandafter\ifx\csname citenamefont\endcsname\relax
  \def\citenamefont#1{#1}\fi
\expandafter\ifx\csname url\endcsname\relax
  \def\url#1{\texttt{#1}}\fi
\expandafter\ifx\csname urlprefix\endcsname\relax\def\urlprefix{URL }\fi
\providecommand{\bibinfo}[2]{#2}
\providecommand{\eprint}[2][]{\url{#2}}

\bibitem[{\citenamefont{{Olinto}}(2000)}]{reviews}
\bibinfo{author}{\bibfnamefont{A.~V.} \bibnamefont{{Olinto}}},
  \bibinfo{journal}{Phys. \ Rep.} \textbf{\bibinfo{volume}{333}},
  \bibinfo{pages}{329} (\bibinfo{year}{2000}). 


\bibitem[{\citenamefont{Greisen}(1966)}]{GZK}
\bibinfo{author}{\bibfnamefont{K.}~\bibnamefont{Greisen}},
  \bibinfo{journal}{Phys. \ Rev. \ Lett.} \textbf{\bibinfo{volume}{16}},
  \bibinfo{pages}{748} (\bibinfo{year}{1966}); 
\bibinfo{author}{\bibfnamefont{G.~T.} \bibnamefont{Zatsepin}} \bibnamefont{and}
  \bibinfo{author}{\bibfnamefont{V.~A.} \bibnamefont{Kuzmin}},
  \bibinfo{journal}{Sov. \ Phys. \ JETP \ Lett.} \textbf{\bibinfo{volume}{4}},
  \bibinfo{pages}{78} (\bibinfo{year}{1966}).

\bibitem[{\citenamefont{Berezinsky et~al.}(1990)\citenamefont{Berezinsky,
  Bulanov, Dogiel, Ginzburg, and Ptuskin}}]{pgamma}
F.~W. Stecker, Phys. Rev. Lett. \textbf{21}, 1016 (1968);
V.~S. Berezinsky and S.~I. Grigorieva, Astron. Astrophys. \textbf{199}, 1 (1988);
\bibinfo{author}{\bibfnamefont{A.}~\bibnamefont{{Mucke}}} {\it et al.},
  \bibinfo{journal}{Publ. \ Astron. \ Soc. \ Aus.}
  \textbf{\bibinfo{volume}{16}}, \bibinfo{pages}{160} (\bibinfo{year}{1999}); 


\bibitem[{\citenamefont{{De Marco} et~al.}(2003)\citenamefont{{De Marco},
  Blasi, and Olinto}}]{daniel}
\bibinfo{author}{\bibfnamefont{D.}~\bibnamefont{{De Marco}}},
  \bibinfo{author}{\bibfnamefont{P.}~\bibnamefont{Blasi}}, \bibnamefont{and}
  \bibinfo{author}{\bibfnamefont{A.~V.} \bibnamefont{Olinto}},
  \bibinfo{journal}{Astropart. \ Phys.} \textbf{\bibinfo{volume}{20}},
  \bibinfo{pages}{53} (\bibinfo{year}{2003}).

\bibitem[{\citenamefont{Dolag et~al.}(2005)\citenamefont{Dolag, Grasso,
  Springel, and Tkachev}}]{klaus}
\bibinfo{author}{\bibfnamefont{K.}~\bibnamefont{Dolag}} {\it et al.},
  \bibinfo{journal}{J. \ Cosmol. \ Astropart. \ Phys.}
  \textbf{\bibinfo{volume}{1}}, \bibinfo{pages}{9} (\bibinfo{year}{2005}).

\bibitem[{\citenamefont{Sigl et~al.}(2004)\citenamefont{Sigl, Miniati, and
  Ensslin}}]{minia}
\bibinfo{author}{\bibfnamefont{G.}~\bibnamefont{Sigl}},
  \bibinfo{author}{\bibfnamefont{F.}~\bibnamefont{Miniati}}, \bibnamefont{and}
  \bibinfo{author}{\bibfnamefont{T.~A.} \bibnamefont{Ensslin}},
  \bibinfo{journal}{Phys. \ Rev. \ D} \textbf{\bibinfo{volume}{70}},
  \bibinfo{pages}{43007} (\bibinfo{year}{2004}).

\bibitem[Bonometto (1971)]{cascade}
S.~A. Bonometto, Lett. Nuovo Cimento \textbf{1}, 677 (1971);
M.~C. Allcock and J. Wdowczyk, Nuovo Cimento \textbf{9B}, 315 (1972).

\bibitem[{\citenamefont{Ferrigno et~al.}(2005)\citenamefont{Ferrigno, Blasi,
  and {De Marco}}}]{carlo}
\bibinfo{author}{\bibfnamefont{C.}~\bibnamefont{Ferrigno}},
  \bibinfo{author}{\bibfnamefont{P.}~\bibnamefont{Blasi}}, \bibnamefont{and}
  \bibinfo{author}{\bibfnamefont{D.}~\bibnamefont{{De Marco}}},
  \bibinfo{journal}{Astropart. \ Phys.} \textbf{\bibinfo{volume}{23}},
  \bibinfo{pages}{211} (\bibinfo{year}{2005}).

\bibitem[{\citenamefont{Aharonian et~al.}(1994)\citenamefont{Aharonian, Coppi,
  and V{\"o}lk}}]{halo}
\bibinfo{author}{\bibfnamefont{F.~A.} \bibnamefont{Aharonian}},
  \bibinfo{author}{\bibfnamefont{P.~S.} \bibnamefont{Coppi}}, \bibnamefont{and}
  \bibinfo{author}{\bibfnamefont{H.~J.} \bibnamefont{V{\"o}lk}},
  \bibinfo{journal}{Astrophys. \ J. \ Lett.} \textbf{\bibinfo{volume}{423}},
  \bibinfo{pages}{5} (\bibinfo{year}{1994}).

\bibitem[{\citenamefont{{Aharonian}}(2002)}]{felixp}
J. Wdowczyk, W. Tkaczyk and A.~W. Wolfendale, J. Phys. A \textbf{5}, 1419 (1972);
\bibinfo{author}{\bibfnamefont{F.~A.} \bibnamefont{{Aharonian}}},
  \bibinfo{journal}{Mon. \ Not. \ R. \ Astron. \ Soc.}
  \textbf{\bibinfo{volume}{332}}, \bibinfo{pages}{215} (\bibinfo{year}{2002}).

\bibitem[{\citenamefont{Gould and Rephaeli}(1978)}]{gould}
F.~W. Stecker, Astrophys. Space Sci. \textbf{20}, 47 (1973);
\bibinfo{author}{\bibfnamefont{R.~J.} \bibnamefont{Gould}} \bibnamefont{and}
  \bibinfo{author}{\bibfnamefont{Y.}~\bibnamefont{Rephaeli}},
  \bibinfo{journal}{Astrophys. \ J.} \textbf{\bibinfo{volume}{225}},
  \bibinfo{pages}{318} (\bibinfo{year}{1978}).

\bibitem[{\citenamefont{{Vall{\' e}e}}(2004)}]{Bfield}
\bibinfo{author}{\bibfnamefont{J.~P.} \bibnamefont{{Vall{\' e}e}}},
  \bibinfo{journal}{New Astronomy Review} \textbf{\bibinfo{volume}{48}},
  \bibinfo{pages}{763} (\bibinfo{year}{2004}). 

\bibitem[{\citenamefont{Clark et~al.}(1970)\citenamefont{Clark, Brown, and
  Alexander}}]{radiobackground}
\bibinfo{author}{\bibfnamefont{T.~A.} \bibnamefont{Clark}},
  \bibinfo{author}{\bibfnamefont{L.~W.} \bibnamefont{Brown}}, \bibnamefont{and}
  \bibinfo{author}{\bibfnamefont{J.~K.} \bibnamefont{Alexander}},
  \bibinfo{journal}{\nat} \textbf{\bibinfo{volume}{228}}, \bibinfo{pages}{847}
  (\bibinfo{year}{1970}).

\bibitem[{\citenamefont{Waxman and Coppi}(1996)}]{deflection}
\bibinfo{author}{\bibfnamefont{E.}~\bibnamefont{Waxman}} \bibnamefont{and}
  \bibinfo{author}{\bibfnamefont{P.}~\bibnamefont{Coppi}},
  \bibinfo{journal}{Astrophys. \ J. \ Lett.} \textbf{\bibinfo{volume}{464}},
  \bibinfo{pages}{75} (\bibinfo{year}{1996}).

\bibitem[{\citenamefont{{Aharonian}}(2003)}]{texas}
www.mpi-hd.mpg.de/hfm/HESS/HESS.html

\bibitem[{\citenamefont{Gould and Schreder}(1966)}]{IRabs}
\bibinfo{author}{\bibfnamefont{R.~J.} \bibnamefont{Gould}} \bibnamefont{and}
  \bibinfo{author}{\bibfnamefont{G.}~\bibnamefont{Schreder}},
  \bibinfo{journal}{Phys. \ Rev. \ Lett.} \textbf{\bibinfo{volume}{16}},
  \bibinfo{pages}{252} (\bibinfo{year}{1966}); 
\bibinfo{author}{\bibfnamefont{J.~R.} \bibnamefont{{Primack}}} {\it et al.},
in \emph{\bibinfo{booktitle}{American Institute
  of Physics Conference Series}} (\bibinfo{year}{2001}), p.
  \bibinfo{pages}{463}.

\bibitem[{\citenamefont{V{\"o}lk et~al.}(1996)\citenamefont{V{\"o}lk,
  Aharonian, and Breitschwerdt}}]{confinement}
\bibinfo{author}{\bibfnamefont{H.~J.} \bibnamefont{V{\"o}lk}},
  \bibinfo{author}{\bibfnamefont{F.~A.} \bibnamefont{Aharonian}},
  \bibnamefont{and}
  \bibinfo{author}{\bibfnamefont{D.}~\bibnamefont{Breitschwerdt}},
  \bibinfo{journal}{Space \ Sci. \ Rev.} \textbf{\bibinfo{volume}{75}},
  \bibinfo{pages}{279} (\bibinfo{year}{1996}); 
\bibinfo{author}{\bibfnamefont{V.~S.} \bibnamefont{Berezinsky}},
  \bibinfo{author}{\bibfnamefont{P.}~\bibnamefont{Blasi}}, \bibnamefont{and}
  \bibinfo{author}{\bibfnamefont{V.~S.} \bibnamefont{Ptuskin}},
  \bibinfo{journal}{Astrophys. \ J.} \textbf{\bibinfo{volume}{487}},
  \bibinfo{pages}{529} (\bibinfo{year}{1997}).


\bibitem[{\citenamefont{Ghisellini and Celotti}(2001)}]{ghisella}
S. Rawlings and R. Saunders, Nature (London) \textbf{349}, 138 (1991).

\bibitem[De Marco (2004)]{daniel2}
P. Blasi and D. De Marco, Astropart. Phys. \textbf{20}, 559 (2004).

\bibitem[{\citenamefont{Waxman}(1995)}]{waxman}
\bibinfo{author}{\bibfnamefont{E.}~\bibnamefont{Waxman}},
  \bibinfo{journal}{Phys. \ Rev. \ Lett.} \textbf{\bibinfo{volume}{75}},
  \bibinfo{pages}{386} (\bibinfo{year}{1995}).

\bibitem[{\citenamefont{Vietri}(1995)}]{mario}
\bibinfo{author}{\bibfnamefont{M.}~\bibnamefont{Vietri}},
  \bibinfo{journal}{Astrophys. \ J.} \textbf{\bibinfo{volume}{453}},
  \bibinfo{pages}{883} (\bibinfo{year}{1995}).

\end{thebibliography}

\end{document}